%% file: Audio-Zero.tex
\documentclass{article} 
\usepackage{iclr2025_conference,times}

\input{math_commands.tex}

\usepackage{url}
\usepackage{bbm}
\usepackage{times}
\usepackage{latexsym}
\usepackage[T1]{fontenc}
\usepackage[utf8]{inputenc}
\usepackage{microtype}
\usepackage{inconsolata}
\usepackage{graphicx}

\usepackage{algorithm}
\usepackage{algpseudocode}
\usepackage{enumitem}
\usepackage{booktabs}
\usepackage{multirow}
\usepackage{array}
\usepackage{amsmath,amssymb}
\usepackage{pifont}
\usepackage{pgfplots}
\usepackage{xcolor}
\usepackage{colortbl} 
\usepackage{url}
\pgfplotsset{compat=1.18}

\usepackage{tcolorbox} 
\tcbuselibrary{skins, breakable} 

\usepackage{tikz}
\usetikzlibrary{arrows.meta, positioning, fit, backgrounds}

\newcommand{\cmark}{\ding{51}}
\newcommand{\xmark}{\ding{55}}

\newcommand{\squishlist}{
   \begin{list}{$\bullet$}
    { \setlength{\itemsep}{0pt}      \setlength{\parsep}{0pt}
      \setlength{\topsep}{-3pt}       \setlength{\partopsep}{0pt}
      \setlength{\listparindent}{-2pt}
      \setlength{\itemindent}{-5pt}
      \setlength{\leftmargin}{1em} \setlength{\labelwidth}{0em}
      \setlength{\labelsep}{0.5em} } }

\newcommand{\squishend}{
    \end{list}  }

\usepackage{array}
\usepackage{multirow}
\usepackage[table]{xcolor}

\usepackage{wrapfig}

\usepackage{authblk}
\usepackage[
  colorlinks=true, 
  citecolor=black,  
  breaklinks=true 
]{hyperref}
\usepackage[symbol]{footmisc}
\title{Audio-Zero: Label-Free Self-Evolution for Fine-Grained Audio Reasoning}

\iclrfinalcopy 
\begin{document}

\author[ ]{\textbf{Siqian Tong\textsuperscript{1 2},\hspace{8pt} Xuan Li\textsuperscript{1 2*},\hspace{8pt} Chaozhuo Li\textsuperscript{3*}, \hspace{8pt} Baolong Bi\textsuperscript{2 4}, \hspace{8pt} Yiwei Wang\textsuperscript{5}}, \\\textbf{Yujun Cai\textsuperscript{6}, \hspace{8pt} Shenghua Liu\textsuperscript{2 4}, \hspace{8pt} Chengpeng Hao\textsuperscript{1 2}}}

\affil[ ]{\textsuperscript{1}Institute of Acoustics, Chinese Academy of Sciences \hspace{3pt} \textsuperscript{2}University of Chinese Academy of Sciences \hspace{3pt} \textsuperscript{3}Beijing Academy of Artificial Intelligence \hspace{3pt} \textsuperscript{4}
Institute of Computing Technology, Chinese
Academy of Sciences \hspace{3pt} \textsuperscript{5}University of California, Merced \hspace{3pt} \textsuperscript{6}The University of Queensland}

\maketitle
{
  \renewcommand{\thefootnote}{$\ast$}
  \footnotetext{Corresponding authors.} 
}
\begin{abstract}
Large Audio Language models (LALMs) have made rapid progress on acoustic understanding, yet they still struggle with fine-grained audio reasoning (e.g., recognizing event order, repetitions and duration). Existing post-training methods heavily rely on expensive external labels or provide only coarse semantic signals. To bridge this gap, we introduce Audio-Zero, the first label-free self-evolution framework in the field of LALMs that improves fine-grained auditory perception and reasoning. Audio-Zero constructs an auditory self-play game from unlabeled audio contrast pairs: most players hear a reference audio, while one odd listener hears a subtle variant. The model first generates clues describing what it hears and then identifies the odd listener by reasoning over inconsistencies among clues. Since the odd listener is known by construction, the game provides verifiable rewards without any annotated answers. Experiments with Qwen2-Audio-7B-Instruct and Qwen2.5-Omni-7B on TREA, MMAU Test-mini and MMAR show that Audio-Zero improves fine-grained audio reasoning while preserving broad audio understanding. Evolutionary and diagnostic analyses further reveal that increasingly fine-grained auditory descriptions emerge naturally from game pressure.
\end{abstract}

\section{Introduction}

Large Audio Language Models (LALMs) have demonstrated strong capability in understanding diverse audio inputs, positioning them as a promising general-purpose paradigm for audio understanding and reasoning~\citep{chu2024qwen2,xu2025qwen3}. 
However, high-quality supervision for LALMs is scarce and hard to scale~\citep{deshmukh2023pengi,tang2024salmonn}, as audio's continuous, temporally extended nature disperses key acoustic cues across long sequences~\citep{mei2024wavcaps}, making fine-grained annotation inherently challenging. 
As a result, popular audio-caption and
audio-QA datasets~\citep{kumar2025mmau,ma2025mmar} rely on loosely-matched, web-scraped descriptions or coarse synthetic text, which capture only general semantics while overlooking detailed temporal-acoustic structure like transient sound events, precise event durations or subtle acoustic pitch changes. 
This bottleneck motivates a central question: \emph{can LALMs self-improve their fine-grained auditory perception and reasoning directly from unlabeled audio, without relying on ground-truth labels, pre-defined QA pairs, or explicit reasoning traces?} 

Self-evolving paradigms, which improve LLM capability without dense human supervision, have been extensively studied in the text and vision domains.
Generally, these paradigms establish autonomous feedback loops by exploiting intrinsic environment constraints or task-inherent rules, effectively transforming self-generated outputs into verifiable optimization signals.
For instance, text-based methods select self-generated reasoning paths based on symbolic correctness \citep{zelikman2022star} or bypass human annotations through zero-sum game dynamics \citep{chen2024self, liu2025spiral, huang2025r, zhao2026absolute}. 
Similarly, Large Vision-Language Models exploit spatially structured, static evidence to construct verifiable visual rewards without manual guidance \citep{he2025visplay, wang2025vision}. 
However, these paradigms fail to directly transfer to the auditory domain. While text relies on discrete symbolic matching and vision allows global, static snapshot scanning, audio is inherently a continuous temporal stream. Crucial acoustic evidence, such as a transient sound event or a subtle ordering difference, unfolds dynamically over time and blends seamlessly into the background signal. Consequently, continuous acoustic cues cannot be verified via exact matching or static inspection, meaning that a label-free audio self-improvement framework demands an entirely audio-specific verification mechanism.
\begin{figure*}[t]
    \centering
    \includegraphics[width=\textwidth]{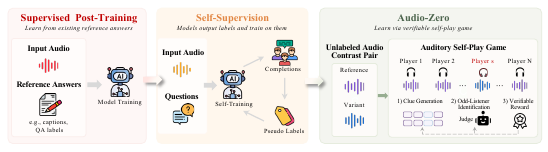}
    \caption{From external supervision to self-evolution. Existing audio-language learning paradigms depend on either external supervision or self-generated supervision. Audio-Zero instead converts unlabeled audio contrast pairs into a verifiable auditory self-play game, where supervision is derived from game outcomes, enabling self-evolution without external labels or reasoning traces.}
    \vspace{-8pt}
    \label{fig:intro}
\end{figure*}

Remaining within the audio domain, existing unsupervised or weakly supervised routes attempt to mitigate this annotation bottleneck, yet they face two fundamental challenges. First, self-supervised representation learning and self-improvement methods driven by pseudo-labels \citep{wang2025self, zhang2025aqa} inherently fall short in fine-grained perception. Because coarse event tags or global acoustic summaries are sufficient to optimize standard training objectives, models rarely possess the incentive to describe details. Second, this lack of precise perception impairs multi-step logical deduction, creating a structural perception-reasoning gap~\citep{yang2025sakura, shih2025can}. Crucially, without a verifiable grounding mechanism, training on these ungrounded descriptions triggers cascading hallucinations and error accumulation, where the policy recursively reinforces its own perceptual mistakes. While reinforcement learning and chain-of-thought distillation can alleviate this gap \citep{fan2025incentivizing, wang2026emotionthinker, yang2026teaching}, they remain strictly bound to labeled QA pairs, teacher rationales, or manually crafted, task-specific rewards. These combined limitations heavily motivate the need for a unified, label-free framework that simultaneously incentivizes fine-grained auditory perception and multi-step reasoning through intrinsic verification.


To bridge this gap, we introduce \textbf{Audio-Zero}, the first label-free self-evolution framework tailored for LALMs. Audio-Zero converts unlabeled paired audios into an auditory self-play game. In each game, most players hear a reference audio, while one odd listener hears a subtle variant. During the listening stage, each player describes its audio in a concise clue. During the attribution stage, the model identifies the odd listener by comparing clues. The odd listener identity is generated automatically by the game construction, providing verifiable rewards without human annotation. Audio-Zero optimizes the model with Group Relative Policy Optimization (GRPO)~\citep{shao2024deepseekmath}, alternating between clue generation and odd-listener identification, so that better clues improve decisions and stronger decision pressure encourages more discriminative clues.

We evaluate Audio-Zero on TREA~\citep{bhattacharya2025benchmarking}, MMAU-Test-mini~\citep{sakshi2024mmau} and MMAR~\citep{ma2025mmar}, covering temporal reasoning, general audio understanding and deep audio reasoning. On Qwen2-Audio-7B-Instruct and Qwen2.5-Omni-7B, Audio-Zero improves temporal reasoning while also improving sound, music and speech categories, indicating that it does not trade off broad auditory competence for task-specific reasoning gains. Our contributions are summarized as follows:
\squishlist
\item We propose Audio-Zero, a label-free self-evolution framework for improving fine-grained auditory perception and reasoning in LALMs without human-written captions, QA labels, or reasoning traces.
\item We design an auditory self-play game that converts unlabeled audio contrast pairs into verifiable training signals by coupling clue generation with odd-listener identification.
\item We conduct comprehensive experiments on TREA, MMAU Test-mini and MMAR using only 2k unlabeled pairs across Qwen2-Audio-7B-Instruct and Qwen2.5-Omni-7B, achieving substantial gains.
\squishend


\section{Audio-Zero: A Label-Free Self-Evolution Framework}

\begin{figure*}[t]
    \centering
    \includegraphics[width=\textwidth]{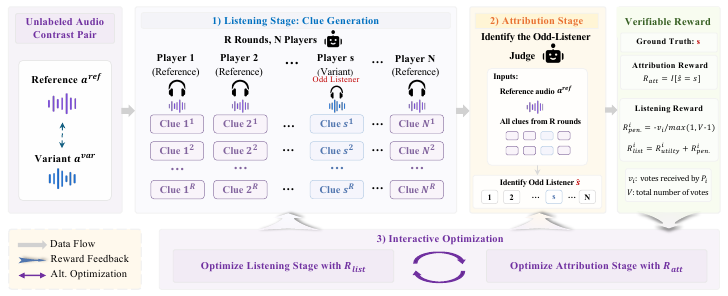}
    \caption{Overview of Audio-Zero. Audio-Zero formulates label-free auditory learning as a multi-round self-play game. Given an unlabeled audio contrast pair, players generate clues from different listening perspectives, while a judge identifies the odd listener. The known game outcome provides verifiable rewards that alternately optimize listening and attribution policies via GRPO, resulting in progressively improved auditory perception and reasoning.}
    \vspace{-8pt}
    \label{fig:framework}
\end{figure*}

Audio-Zero is a self-evolution training framework that improves fine-grained auditory perception and reasoning in LALMs without relying on any external ground-truth labels. The core idea is to convert unlabeled audio contrast pairs into a multi-player identification game, where the model must first verbalize what it hears and then reason across descriptions to find the odd listener. The odd listener's identity is procedurally assigned during game construction, serving as the intrinsic ground truth to yield an inherently verifiable reward embedded within the game structure. Rather than relying on static labels, we set up a tight feedback loop where clue generation and logical attribution mutually reinforce each other, creating an autonomous evolutionary trajectory as illustrated in Figure~\ref{fig:framework}.

\subsection{Problem Definition}

Let $\pi_\theta$ denote a pre-trained LALM parameterized by $\theta$. We assume access to a training collection of \emph{unlabeled audio contrast pairs} $\mathcal{D} = \{(a^{\mathrm{ref}}_k,\, a^{\mathrm{var}}_k)\}_{k=1}^{K}$, where each pair shares high-level
semantics but exhibits subtle acoustic differences such as differing event orders, repetition counts, durations, missing sounds, or background scenes. No human transcripts, QA annotations, or reasoning traces are required.

For each game instance, a single model $\pi_\theta$ sequentially
plays $N$ distinct player roles. An odd listener identity $s \in \{1,\ldots,N\}$ is procedurally assigned: the model acting as player $s$ receives the variant audio
$a^{\mathrm{var}}$ as its private input, while all other players $i \neq s$ receive $a^{\mathrm{ref}}$. Over $R$ listening rounds, the model generates a clue for each
player role given its private audio and the public history $H^r$ (all clues generated up to round $r-1$). In the attribution stage, the model takes an judge perspective, reading the full clue history $H^{R+1}$ together with
$a^{\mathrm{ref}}$ to predict the odd listener $\hat{s} \in \{1,\ldots,N\}$. Our goal is to optimize the model parameters $\theta$ via GRPO by maximizing game performance across both stages. Specifically, we employ a dual-reward design that jointly reinforces correct odd-listener attribution and penalizes uninformative clues. By alternating between these twin objectives, we aim to find optimal parameters $\theta^*$ that establish a cooperative feedback loop between clue generation and logical identification, thereby eliciting stronger fine-grained auditory perception and reasoning than the base policy.

\subsection{Auditory Self-Play Game}

Each training step draws a contrast pair $(a^{\mathrm{ref}}, a^{\mathrm{var}})$
and instantiates a game with $N$ virtual players.The odd listener
$s \sim \mathrm{Uniform}(\{1,\ldots,N\})$ is sampled uniformly and kept
hidden from the model; it is known only to the training environment.
Player $i$ receives audio
\begin{equation}
    a_i =
    \begin{cases}a^{\mathrm{var}}, & i = s\\
        a^{\mathrm{ref}}, & i \neq s
    \end{cases}
\end{equation}
The game proceeds in two coupled stages.

\paragraph{Listening Stage.}

In the listening stage, each player $i \in \{1,\ldots,N\}$ generates a
natural-language clue $z_i^r$ in each round $r \in \{1,\ldots,R\}$
by conditioning on its own audio and the shared public history:
\begin{equation}
    z_i^r \sim \pi_\theta\!\left(\,\cdot \mid a_i,\, H_i^r\right).
\end{equation}
A well-formed description is expected to communicate concrete auditory
observations, such as sound event types, temporal ordering, repetition counts,
acoustic preperties and scene context, while refraining from
directly disclosing whether the speaker holds the reference or variant
audio.
This constraint forces the model to extract fine-grained perceptual
detail from the audio, organize it into informative natural language,
and reason strategically about what to reveal given the public history.
After all $N \times R$ descriptions have been generated, the full set
$Z = \{z_i^r\}_{i \in [N],\, r \in [R]}$ is assembled for the
attribution stage.

\paragraph{Attribution Stage.}

In the attribution stage, the model acts as a judge.
Given the reference audio $a^{\mathrm{ref}}$ and the complete description
set $Z$, it predicts the odd listener:
\begin{equation}
    \hat{s} \sim \pi_\theta\!\left(\,\cdot \mid a^{\mathrm{ref}},\, Z\right)
\end{equation}
This requires higher-order reasoning: the model must compare descriptions
across players, detect semantic inconsistencies (e.g., one player reporting
a knock \emph{before} speech while others report the reverse) and weigh
whether such inconsistencies reflect genuine perceptual differences or
merely vague phrasing.

This coupling between the two stages is architecturally central to Audio-Zero. The listening stage produces the
evidence; the attribution stage verifies whether that evidence is
discriminative enough to solve the game.

\subsection{Label-Free Reward and Self-Evolution Objective}

The two stages of Audio-Zero are each associated with a dedicated reward signal. Both are derived entirely from game mechanics and require no external annotation.

\paragraph{Attribution Reward.}
The attribution reward measures whether the model, acting as the judge, correctly identifies the odd listener from the full description history.

It is defined as the binary game outcome:
\begin{equation}
    R_{\mathrm{att}} = \mathbb{I}[\hat{s} = s]
\end{equation}
The ground truth $s$ is generated by the environment at game construction time and is never
provided to the model as input.  A correct identification yields a
reward of $1$; any incorrect prediction yields $0$.

\paragraph{Listening Reward.}
The listening reward measures whether the model captures accurate,
fine-grained auditory details and expresses them in descriptions that
are strategically effective within the game. To achieve this, the reward couples a content score with a dynamic game-theoretic penalty.

\squishlist
\item \textbf{Content utility} rewards clues that are (i) grounded in specific
auditory evidence (sound events, temporal relations, acoustic attributes),
(ii) non-repetitive across rounds and (iii) free from identity leakage
(explicit statements that the speaker is or is not the odd listener).
This component is computed via a rule-based scorer that checks the
presence of audio-descriptive terms and penalizes verbatim repetitions
and banned phrases.

\item \textbf{Vote-aware penalty} adjusts the reward based on the
judge's attribution outcomes.
Let $v_i$ denote the number of attribution votes received by player $i$ within the game, and let $V$ be the total number of votes.
If the odd player $s$ generates descriptions that are trivially
self-revealing and attract disproportionate votes, its reward is
penalized; equally, a non-odd player whose descriptions draw unwarranted
suspicion incurs the same penalty.
Formally, the listening reward for player $i$ is
\begin{equation}
    R_{\mathrm{list}}^{i} =
    R_{\mathrm{utility}}^{i}
    - \lambda \cdot \frac{v_i}{\max(1,\, V - 1)},
\end{equation}
where $\lambda > 0$ controls the
penalty strength. This makes description generation strategic:
a good description should be grounded, informative and useful for exposing the true odd listener rather than drawing unwarranted suspicion to oneself.
\squishend

During alternating optimization, the active reward is defined as:
\begin{equation}
R =
\begin{cases}
R_{\mathrm{list}}^{i}, & \text{listening stage for player } i\\
R_{\mathrm{att}}, & \text{attribution stage}
\end{cases}
\end{equation}
This separation prevents gradients from the inactive stage from interfering with the current update.
\subsection{Interactive Optimization}

Audio-Zero uses Group Relative Policy Optimization~(GRPO)~\citep{shao2024deepseekmath}
as its underlying RL algorithm.
For a given prompt $x$ (audio input plus game context), GRPO samples a
group of $G$ completions $\{y_j\}_{j=1}^{G}$ from the old policy
$\pi_{\theta_{\mathrm{old}}}(\cdot \mid x)$, assigns rewards
$\{R_j\}_{j=1}^{G}$, and computes group-normalized advantages:
\begin{equation}
    A_j =
    \frac{R_j - \mathrm{mean}(\{R_j\}_{j=1}^{G})}
    {\mathrm{std}(\{R_j\}_{j=1}^{G}) + \varepsilon}.
\end{equation}
The policy update maximizes the clipped surrogate objective with a
KL-divergence penalty against a frozen reference policy $\pi_{\mathrm{ref}}$:
\begin{equation}
{\small
\begin{aligned}
    \mathcal{L}(\theta) =\, &
    \mathbb{E}_{\substack{x \sim \mathcal{D} \\
    \{y_j\}_{j=1}^{G} \sim \pi_{\theta_{\mathrm{old}}}(\cdot \mid x)}}
    \Bigg[
    \frac{1}{G} \sum_{j=1}^{G}
    \min \Big(
        r_j(\theta) A_j, \mathrm{clip}(r_j(\theta), 1\pm \epsilon) A_j
    \Big)
    \Bigg]- \beta
    D_{\mathrm{KL}}(\pi_\theta \| \pi_{\mathrm{ref}}).
\end{aligned}
}
\label{eq:grpo_objective}
\end{equation}
where
\begin{equation}
    r_j(\theta)=
    \frac{\pi_\theta(y_j \mid x)}
    {\pi_{\theta_{\mathrm{old}}}(y_j \mid x)}
\end{equation}
is the importance sampling ratio.

Here, $j$ indexes sampled completions within the GRPO group, while $i$
indexes the player role in the auditory self-play game.
The prompt $x$, completion $y_j$, and reward $R_j$ take different forms
in each phase:
\squishlist
\item \textbf{List. phase}: $x=(a_i,H_i^r)$; $y_j$ is a sampled clue
candidate for player $i$ under the current listening context and
$R_j=R_{\mathrm{list}}^{i}(y_j)$.
\item \textbf{Attr. phase}: $x=(a^{\mathrm{ref}},Z)$; $y_j$ is a sampled
odd-listener prediction and $R_j=R_{\mathrm{att}}(y_j)$.
\squishend

\paragraph{Alternating optimization.}
Training alternates between two phases at every update step.
In the attribution phase, the clues produced by the current policy are
treated as fixed context. The model is optimized solely on
$R_{\mathrm{att}}$, learning to identify the odd listener from existing
evidence.
In the listening phase, the model is optimized solely on
$R_{\mathrm{list}}^{(i)}$, with attribution outcomes serving as
vote-aware feedback for player-specific clue generation.

This alternation creates a self-reinforcement loop. Stronger attribution
raises the bar for clue quality, because only genuinely discriminative
clues lead to correct identifications. In turn, higher-quality clues
provide richer evidence, making it easier to train attribution further.
Neither stage reaches a fixed point in isolation; they co-evolve
throughout training.


\section{Experiments and Analysis}
\label{sec:experiments}

\subsection{Experimental Setup}
\label{sec:setup}

\paragraph{Training data.}
We instantiate Audio-Zero with an existing paired audio preference dataset Audio-alpaca~\citep{majumder2024tango}.
Each item contains a chosen audio and a rejected audio that share prompt-level semantics but may differ in event order, missing events, repetitions, or acoustic quality.
We sample and filter 2{,}000 pairs from the full collection for
post-training.
The dataset serves as a practical source of unlabeled auditory contrast;
the choice of collection is an experimental instantiation rather than
a methodological requirement.

\paragraph{Models.}
We apply Audio-Zero to two base models: Qwen2-Audio-7B-Instruct~\citep{chu2024qwen2},
a strong audio-language model and Qwen2.5-Omni-7B~\citep{xu2025qwen25omnitechnicalreport},
a unified omni-modal model.

\paragraph{Evaluation Benchmarks.}
We evaluate model performance on three benchmarks using
Multiple-Choice Question (MCQ) accuracy on two general
audio reasoning benchmarks and one fine-grained benchmark
closely related to the training task.
MMAU Test-mini~\citep{sakshi2024mmau} is a 1{,}000-question benchmark for general audio understanding and reasoning, covering Sound, Music and Speech.
MMAR~\citep{ma2025mmar} evaluates deep reasoning over speech, audio, music,
and mixed-modality audio.
TREA~\citep{bhattacharya2025benchmarking} targets temporal reasoning with three subsets:
Order, Count and Duration (200 questions each).

\paragraph{Implementation details.}
Unless otherwise specified, we use $N=4$ players and $R=2$ listening rounds per game.
Models are optimized with GRPO using $G=6$ completions per prompt,
learning rate $1\times10^{-5}$, bf16 precision and KL regularization
coefficient $\beta=0.04$.
Additional hyperparameters are in Appendix~\ref{app:details}.

\subsection{Main Results: Reasoning Improves without Hurting General Audio Ability}


\begin{table*}[t]
\centering
\setlength{\tabcolsep}{3.5pt}
\resizebox{\textwidth}{!}{%
\begin{tabular}{l | cccc | cccc | cccc}
\toprule
\multirow{2}{*}{\textbf{Method}} & \multicolumn{4}{c|}{\textbf{MMAU Test-mini}} & \multicolumn{4}{c|}{\textbf{MMAR}} & \multicolumn{4}{c}{\textbf{TREA}} \\
\cmidrule(lr){2-5} \cmidrule(lr){6-9} \cmidrule(lr){10-13}
 & \textbf{Sound} & \textbf{Music} & \textbf{Speech} & \textbf{Avg.} & \textbf{Sound} & \textbf{Music} & \textbf{Speech} & \textbf{Avg.} & \textbf{Order} & \textbf{Count} & \textbf{Dur.} & \textbf{Avg.} \\
\midrule

\rowcolor{blue!5}
\multicolumn{13}{c}{\textbf{Qwen2-Audio-7B-Instruct}} \\
Base Policy & 60.96 & 58.68 & 52.85 & 57.50 & 46.67 & 31.55 & 40.82 & 41.20 & 59.00 & 27.50 & 35.50 & 40.67 \\

\midrule

\multicolumn{13}{l}{\cellcolor[HTML]{F8F8F8}\textit{Label-Dependent}} \\
R1-AQA~\citep{li2025reinforcement} & 63.06 & 59.28 & 54.65 & 59.00 & 47.27 & 34.95 & 43.54 & 44.50 & 62.50 & 28.50 & 43.00 & 44.67 \\
Audio-Thinker~\citep{wu2026audio}
& \underline{63.66} & \underline{58.69} & \underline{56.46} & \underline{59.60}
& \underline{47.88} & \underline{38.35} & \underline{46.26} & \underline{46.80}
& \underline{64.50} & \underline{29.50} & \underline{45.00} & \underline{46.33} \\

\midrule

\multicolumn{13}{l}{\cellcolor[HTML]{F8F8F8}\textit{Label-Free}} \\
AQA-TTRL~\citep{zhang2025aqa} & 61.56 & 58.98 & 53.75 & 58.10 & 46.67 & 33.01 & 42.18 & 42.20 & 59.50 & 28.00 & 37.00 & 41.50 \\
Audio-CoT~\citep{ma2025audio} & 61.26 & 58.38 & 53.75 & 57.80 & 46.06 & 31.55 & 41.50 & 41.50 & 57.50 & 26.00 & 34.00 & 39.17 \\
\rowcolor[HTML]{EFEFEF}
\textbf{Audio-Zero (Ours)} & \textbf{64.56} & \textbf{58.69} & \textbf{58.36} & \textbf{60.50} & \textbf{48.48} & \textbf{41.26} & \textbf{47.96} & \textbf{49.10} & \textbf{66.50} & \textbf{30.00} & \textbf{47.50} & \textbf{48.00} \\

\midrule

\rowcolor{blue!5}
\multicolumn{13}{c}{\textbf{Qwen2.5-Omni-7B}} \\
Base Policy & 70.57 & 68.26 & 63.06 & 68.30 & 56.36 & 47.09 & 59.86 & 56.20 & 90.50 & 58.50 & 61.00 & 70.00 \\

\midrule

\multicolumn{13}{l}{\cellcolor[HTML]{F8F8F8}\textit{Label-Dependent}} \\
R1-AQA~\citep{li2025reinforcement} & 76.58 & 70.66 & 70.27 & 72.50 & 63.64 & 50.49 & 60.54 & 60.80 & 92.00 & 61.00 & 63.50 & 72.17 \\
Audio-Thinker~\citep{wu2026audio}
& \underline{79.28} & \underline{71.56} & \underline{72.97} & \underline{74.60}
& \underline{66.67} & \underline{53.40} & \underline{61.90} & \underline{63.70}
& \underline{92.50} & \underline{63.00} & \underline{64.00} & \underline{73.17}  \\

\midrule

\multicolumn{13}{l}{\cellcolor[HTML]{F8F8F8}\textit{Label-Free}} \\
AQA-TTRL~\citep{zhang2025aqa} & 72.67 & 69.46 & 66.67 & 69.60 & 59.39 & 48.54 & 60.20 & 57.50 & 91.00 & 59.50 & 62.00 & 70.83 \\
Audio-CoT~\citep{ma2025audio} & 71.47 & 68.86 & 66.37 & 68.90 & 57.58 & 47.57 & 59.86 & 56.80 & 89.00 & 57.00 & 59.00 & 68.33 \\
\rowcolor[HTML]{EFEFEF}
\textbf{Audio-Zero (Ours)} & \textbf{81.68} & \textbf{72.75} & \textbf{74.47} & \textbf{76.30} & \textbf{69.70} & \textbf{55.83} & \textbf{60.54} & \textbf{66.20} & \textbf{93.00} & \textbf{64.50} & \textbf{64.50} & \textbf{74.00} \\
\bottomrule
\end{tabular}%
}
\caption{Comprehensive evaluation across multi-modal audio benchmarks. Bold indicates the best performance for each base model and underline represents the best among all baseline methods.}
\label{tab:main_results}
\end{table*}

To comprehensively evaluate Audio-Zero, we compare it against representative baselines spanning different learning paradigms. For the label-dependent group, we evaluate \textit{R1-AQA} \citep{li2025reinforcement} and \textit{Audio-Thinker} \citep{wu2026audio}, which represent strong reinforcement learning and explicit reasoning baselines optimized via ground-truth answers. For the label-free group, we compare against (1) \textit{AQA-TTRL} \citep{zhang2025aqa}, a test-time reinforcement learning paradigm that dynamically optimizes the model's policy via majority-consensus pseudo-label feedback and (2) \textit{Audio-CoT} \citep{ma2025audio}, a training-free prompt-based method encouraging intermediate reasoning. We evaluate all methods across two foundation backbones: Qwen2-Audio-7B-Instruct and Qwen2.5-Omni-7B.

As reported in Table~\ref{tab:main_results}, Audio-Zero consistently achieves the best overall performance across all three benchmarks and both backbones, outperforming not only all label-free test-time approaches but also the strong label-dependent training baselines. Specifically, using merely 2k unlabeled contrast pairs, Audio-Zero reaches an average accuracy of 60.50\% on MMAU Test-mini and 49.10\% on MMAR under Qwen2-Audio-7B-Instruct, outperforming the base policy by 3.00\% and 7.90\% respectively. On the stronger Qwen2.5-Omni-7B backbone, Audio-Zero further yields substantial gains, boosting MMAU Avg. by 8.00\% and MMAR Avg. by 10.00\%. This comprehensive improvement confirms that converting unlabeled audio contrast pairs into an interactive self-play game can unlock latent perception capabilities, boosting both general multi-modal understanding and specialized audio reasoning without any external supervision.

Crucially, the performance gains are particularly pronounced on the TREA subsets, which explicitly target fine-grained, multi-step temporal reasoning (e.g., reaching 66.50\% on Order and 47.50\% on Duration under Qwen2-Audio-7B-Instruct). This dual improvement across both general and fine-grained benchmarks directly reflects the efficacy of our dual-reward framework. While standard label-free methods struggle with vague feedback, Audio-Zero establishes a dynamic game-theoretic constraint between the two optimization stages. Within this loop, the attribution stage acts as a judge that evaluates game outcomes, which inherently forces the listening stage to produce more detailed, grounded clues. Consequently, the interactive competition inherently drives the model to capture and process more fine-grained acoustic information, validating the immense potential of self-evolutionary training.


\subsection{Ablation Study}
\label{sec:ablation}

The core design of Audio-Zero aims to achieve a joint self-evolution of both acoustic perception and multi-step reasoning. To verify whether this evolutionary loop requires the synergy of all components, we evaluate five architectural variants across both Qwen2-Audio-7B-Instruct and Qwen2.5-Omni-7B: (1) \textit{Vanilla GRPO w/o game}, a standard single-stage RL baseline; (2) \textit{Listening-stage only} and (3) \textit{Attribution-stage only}, which restrict optimization to a single stage; (4) \textit{w/o Vote-aware feedback}, which removes the collective voting process; and (5) \textit{Sequential training}, which optimizes the two stages in a static, one-way sequence.

As shown in Table~\ref{tab:ablation}, removing any single component leads to a noticeable performance decline. Isolating either the listening or attribution stage yields marginal improvements, showing that single-role updates cannot sustain a continuous self-evolution loop. Crucially, disabling the vote-aware feedback causes a distinct performance drop on the TREA benchmark (from 48.00\% to 45.50\% on Qwen2; 74.00\% to 72.67\% on Qwen2.5). This empirical drop clearly verifies that collective voting is indispensable, as it suppresses sporadic hallucinations and guides the model with more reliable feedback. Furthermore, the performance gap between \textit{Sequential training} and the full framework underscores that alternating optimization serves as the cornerstone of our framework. Instead of leading to training stagnation, this dynamic alternation drives the co-evolution of both roles: as the two stages update alternately, they continuously push each other forward, ultimately forcing the listening model to deliver increasingly precise and fine-grained acoustic facts in its reasoning steps.

\begin{table*}[t]
\centering
\small
\setlength{\tabcolsep}{2.5pt} 

\resizebox{\textwidth}{!}{
\begin{tabular}{l ccc ccc ccc}
\toprule
 & \multicolumn{3}{c}{\textbf{Components}} & \multicolumn{3}{c}{\textbf{Qwen2-Audio-7B-Instruct}} & \multicolumn{3}{c}{\textbf{Qwen2.5-Omni-7B}} \\
\cmidrule(lr){2-4} \cmidrule(lr){5-7} \cmidrule(lr){8-10}
\textbf{Variant} & \textbf{List.} & \textbf{Attr.} & \textbf{Inter.} & \textbf{TREA} & \textbf{MMAU Test-mini} & \textbf{MMAR} & \textbf{TREA} & \textbf{MMAU Test-mini} & \textbf{MMAR} \\
\midrule
Vanilla GRPO (w/o game)       & \xmark & \xmark & \xmark & 44.67 & 59.00 & 44.50 & 72.17 & 72.50 & 60.80 \\
Listening-stage only          & \cmark & \xmark & \xmark & 45.33 & 59.40 & 45.80 & 72.50 & 73.30 & 61.90 \\
Attribution-stage only        & \xmark & \cmark & \xmark & 45.83 & 59.10 & 47.20 & 72.83 & 72.70 & 63.10 \\
w/o Vote-aware feedback       & \cmark & \cmark & \cmark & 45.50 & 59.20 & 46.80 & 72.67 & 72.90 & 62.70 \\
Sequential training & \cmark & \cmark & \xmark & 46.83 & 60.10 & 47.90 & 73.33 & 74.80 & 64.50 \\
\rowcolor[HTML]{EFEFEF}
\textbf{Audio-Zero (Full)}            & \cmark & \cmark & \cmark & \textbf{48.00} & \textbf{60.50} & \textbf{49.10} & \textbf{74.00} & \textbf{76.30} & \textbf{66.20} \\
\bottomrule
\end{tabular}
}
\caption{Ablation studies on both Qwen2-Audio-7B-Instruct and Qwen2.5-Omni-7B-Instruct backbones across downstream benchmarks. List., Attr. and Inter. denote the Listening-stage, Attribution-stage and Vote-aware Interaction feedback, respectively.}
\label{tab:ablation}
\end{table*}

\subsection{Self-Evolution Dynamics}
We further analyze whether Audio-Zero exhibits self-evolution during training.
At fixed evaluation intervals from 0 to 100 iterations, the active policy
plays self-play games against the frozen initial policy, alternating
between civilian-listener and odd-listener role assignments.
We track game win rates, average listening-stage clue length and
downstream performance on external audio reasoning benchmarks.

\begin{figure*}[t]
\centering
\definecolor{navyblue}{HTML}{1F77B4}
\definecolor{coralorange}{HTML}{FF7F0E}
\definecolor{emeraldgreen}{HTML}{2CA02C}

\begin{tikzpicture}
\begin{axis}[
    width=0.33\linewidth, height=4.3cm, 
    title={(a) Game Win Rate}, xlabel={Iteration}, ylabel={Win Rate (\%)},
    ymin=41, ymax=76, 
    legend style={font=\tiny, at={(0.98,0.02)}, anchor=south east, draw=none, fill=white, fill opacity=0.6, text opacity=1, inner sep=1.5pt, row sep=-3pt},
    xtick={0,20,40,60,80,100}, grid=both, grid style={dashed, gray!30},
    title style={font=\small\bfseries}, label style={font=\small}, tick label style={font=\footnotesize}]
\addplot+[thick, color=navyblue, mark=*, mark size=1.0pt] coordinates {
    (0,50.0) (10,52.0) (20,54.0) (30,52.0) (40,56.0) (50,56.0) (60,64.0) (70,66.0) (80,68.0) (90,70.0) (100,72.0)
};
\addlegendentry{Civilian Listener}
\addplot+[thick, color=coralorange, mark=square*, mark size=1.0pt] coordinates {
    (0,50.0) (10,50.0) (20,52.0) (30,56.0) (40,52.0) (50,58.0) (60,54.0) (70,59.0) (80,64.0) (90,66.0) (100,68.0)
};
\addlegendentry{Odd Listener}
\end{axis}
\end{tikzpicture}
\hskip 0.05cm 
\begin{tikzpicture}
\begin{axis}[
    width=0.33\linewidth, height=4.3cm, 
    title={(b) Clue Completion Length}, xlabel={Iteration}, ylabel={Tokens},
    ymin=120, ymax=450, xtick={0,20,40,60,80,100}, grid=both, grid style={dashed, gray!30},
    title style={font=\small\bfseries}, label style={font=\small}, tick label style={font=\footnotesize}]
\addplot+[thick, color=navyblue, mark=*, mark size=1.0pt] coordinates {
    (1,170.1) (10,148.2) (20,179.4) (30,227.4) (40,243.8) (50,301.3) (60,278.5) (70,329.8) (80,398.8) (90,418.5) (100,423)
};
\end{axis}
\end{tikzpicture}
\hskip 0.05cm
\begin{tikzpicture}
\begin{axis}[
    width=0.33\linewidth, height=4.3cm, 
    title={(c) Audio Reasoning}, xlabel={Iteration}, ylabel={Accuracy (\%)},
    ymin=35, ymax=64,
    legend style={font=\tiny, at={(1.00,-0.00)}, anchor=south east, draw=none, fill=white, fill opacity=0.6, text opacity=1, inner sep=1pt, row sep=-3.5pt},
    xtick={0,20,40,60,80,100}, grid=both, grid style={dashed, gray!30},
    title style={font=\small\bfseries}, label style={font=\small}, tick label style={font=\footnotesize}]
\addplot+[thick, color=navyblue, mark=*, mark size=1.0pt] coordinates {
    (0,40.67) (10,42.10) (20,43.17) (30,43.0) (40,43.33) (50,44.17) (60,43.83) (70,46.0) (80,47.67) (90,46.33) (100,48.00)
};
\addlegendentry{TREA}
\addplot+[thick, color=coralorange, mark=square*, mark size=1.0pt] coordinates {
    (0,41.20) (10,42.50) (20,44.10) (30,44.6) (40,43.60) (50,45.10) (60,46.50) (70,47.30) (80,47.80) (90,48.40) (100,49.10)
};
\addlegendentry{MMAR}
\addplot+[thick, color=emeraldgreen, mark=triangle*, mark size=1.0pt] coordinates {
    (0,57.50) (10,58.20) (20,58.90) (30,58.60) (40,58.40) (50,59.00) (60,59.10) (70,59.60) (80,59.60) (90,60.20) (100,60.50)
};
\addlegendentry{MMAU Test-mini}
\end{axis}
\end{tikzpicture}
\caption{Analysis of self-evolution dynamics across training iterations. Subfigures depict (a) asymmetric game win rates for distinct role assignments, (b) average lengths of generated audio clues and (c) mathematical transfer performance on downstream audio benchmarks.}
\label{fig:evolution}
\end{figure*}
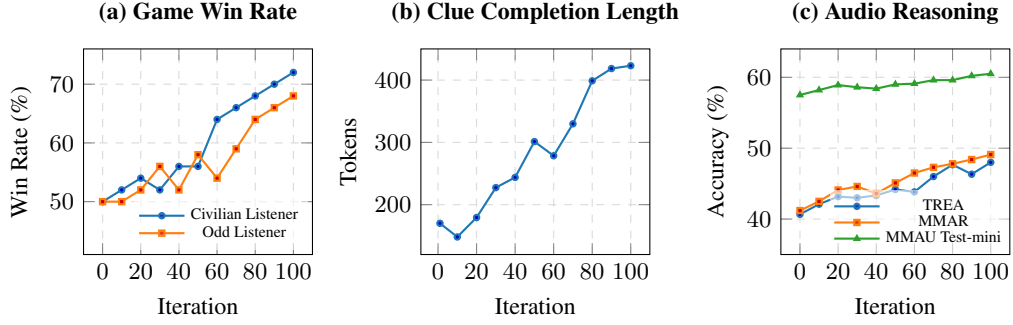

Figure~\ref{fig:evolution} illustrates three trends.
First, game win rates generally increase for both role assignments,
although with local fluctuations, indicating that the model's ability
to identify subtle auditory differences and reason over generated evidence improves throughout training.
Second, listening-stage clue length increases in the early phase and then
stabilizes; its fluctuations indicate that the policy balances informativeness with game effectiveness rather than simply producing longer descriptions.
Crucially, TREA, MMAR and MMAU Test-mini all show an overall upward trend,
with larger gains on the more fine-grained benchmarks, suggesting that
Audio-Zero improves both auditory detail perception and reasoning over
consistency across player descriptions.
Together, these results indicate that the self-play game drives
progressive self-evolution and that the learned capabilities generalize
to external audio reasoning benchmarks.

\subsection{Emergence of Fine-Grained Auditory Descriptions}
\label{sec:diagnostic}

We analyze whether the game pressure
in Audio-Zero drives the model to capture fine-grained auditory details, rather
than merely improving final benchmark accuracy.
We track five metrics over training iterations: \textit{Distinct Event Count}, \textit{Temporal Relation Count}, \textit{Quantity Expression Frequency}, \textit{Acoustic Attribute Density} and \textit{Vagueness Rate}; detailed definitions and matching rules are provided in
Appendix~\ref{app:diagnostic_metrics}.

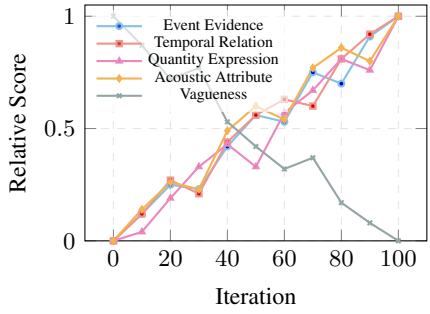
\begin{figure}[h] 
\centering
\begin{minipage}[b]{0.46\linewidth} 
    \centering
    \begin{tikzpicture}[trim left=(current axis.west), trim right=(current axis.east)]
    \definecolor{pastelred}{HTML}{F1948A}
    \definecolor{pastelblue}{HTML}{85C1E9}
    \definecolor{softpink}{HTML}{EC87C0}
    \definecolor{lightorange}{HTML}{F4B350}
    \definecolor{softgray}{HTML}{95A5A6}
    \begin{axis} [
        width=0.95\linewidth, 
        height=4.7cm,        
        ymin=0, ymax=1.05,
        xtick={0,20,40,60,80,100},
        xlabel={Iteration},
        xlabel style={yshift=-0.05cm}, 
        ylabel={Relative Score},
        ylabel style={yshift=0cm},
        legend style={
            font=\tiny,
            at={(0.02,0.98)},
            anchor=north west,
            draw=none,
            fill=white,
            fill opacity=0.65,
            text opacity=1,
            inner sep=1.2pt,
            row sep=-3pt
        },
        grid=both,
        grid style={dashed, gray!20},
        title style={font=\small\bfseries},
        label style={font=\small},
        tick label style={font=\footnotesize}
    ]
    \addplot+[thick, color=pastelblue, mark=*, mark size=1.1pt] coordinates {
        (0,0.00) (10,0.12) (20,0.25) (30,0.23) (40,0.42)
        (50,0.56) (60,0.53) (70,0.75) (80,0.70) (90,0.91) (100,1.00)
    };
    \addlegendentry{Event Evidence}
    \addplot+[thick, color=pastelred, mark=square*, mark size=1.0pt] coordinates {
        (0,0.00) (10,0.12) (20,0.27) (30,0.21) (40,0.44)
        (50,0.56) (60,0.63) (70,0.60) (80,0.81) (90,0.92) (100,1.00)
    };
    \addlegendentry{Temporal Relation}
    \addplot+[thick, color=softpink, mark=triangle*, mark size=1.1pt] coordinates {
        (0,0.00) (10,0.04) (20,0.19) (30,0.33) (40,0.43)
        (50,0.33) (60,0.57) (70,0.67) (80,0.81) (90,0.76) (100,1.00)
    };
    \addlegendentry{Quantity Expression}
    \addplot+[thick, color=lightorange, mark=diamond*, mark size=1.1pt] coordinates {
        (0,0.00) (10,0.14) (20,0.26) (30,0.23) (40,0.49)
        (50,0.60) (60,0.54) (70,0.77) (80,0.86) (90,0.80) (100,1.00)
    };
    \addlegendentry{Acoustic Attribute}
    \addplot+[thick, color=softgray, mark=x, mark size=1.2pt] coordinates {
        (0,1.00) (10,0.87) (20,0.72) (30,0.77) (40,0.53)
        (50,0.42) (60,0.32) (70,0.37) (80,0.17) (90,0.08) (100,0.00)
    };
    \addlegendentry{Vagueness}   
    \end{axis}
    \end{tikzpicture}
    \caption{Diagnostic analysis of generated listening-stage descriptions across
training iterations. All metrics are independently min-max scaled to
$[0,1]$ for visualization after length normalization where applicable.
Higher values indicate stronger fine-grained auditory evidence, except
for Vagueness, where lower values indicate better specificity.} 
    \label{fig:diagnostic}
\end{minipage}\hfill 
\begin{minipage}[b]{0.5\linewidth}
As shown in Figure~\ref{fig:diagnostic}, all detail-related metrics
generally increase during training, while Vagueness consistently
decreases.
Event Evidence and Temporal Relation scores indicate that the model
gradually moves from generic audio summaries toward descriptions that
mention concrete sound events and organize them into temporal structures.
The growth of Quantity Expression suggests improved sensitivity to
repetition and counting-related details, while the increase in Acoustic Attribute indicates that the model becomes more attentive to properties such as loudness, distance, clarity and acoustic texture.
Meanwhile, the decline of Vagueness indicates that the model relies less
on underspecified phrases such as ``some sounds'' or ``various noises.''

Since these metrics are independently scaled, we focus on their evolutionary trajectories during training rather than cross-metric absolute values. Overall, the results show that Audio-Zero's self-play game does not merely encourage longer outputs.
    \vspace{-0.3cm} 
\end{minipage}
\end{figure}
 Instead, attribution pressure pushes the model toward descriptions that are more event-rich, temporally grounded, acoustically detailed and discriminative across players, suggesting that fine-grained auditory detail perception emerges through the self-evolution process.

\subsection{Case Study}
Table~\ref{tab:case_study_table} presents a qualitative case study of
Audio-Zero's self-evolution on a temporal-reasoning case.
At Stage 1, the base policy produces coarse and underspecified
descriptions, so the attribution stage can only rely on surface event
overlap and fails to identify the odd listener.
By Stage 2, the model begins to describe the basic event sequence and
the attribution stage can detect a local order mismatch between player
descriptions, although the reasoning remains relatively shallow.
At Stage 3, Audio-Zero generates more fine-grained auditory evidence,
including event order, repetition count, acoustic strength and
background context.
The attribution stage also adopts a fine-grained cross-checking mechanism: it compares the majority timeline against each player's description, localizes the
temporal inconsistency and correctly identifies the odd listener.
This case illustrates that Audio-Zero improves both listening-stage
detail perception and attribution-stage reasoning through self-evolution.

\begin{table*}[t]
\centering
\scriptsize
\setlength{\tabcolsep}{6pt}
\definecolor{boxred}{HTML}{FDEDEC}
\definecolor{boxyellow}{HTML}{FEF9E7}
\definecolor{boxgreen}{HTML}{EAF2F8}

\begin{tabular}{p{0.18\textwidth} | p{0.78\textwidth}}
\toprule
\multicolumn{2}{c}{\textbf{Case Study: Behavioral Emergence on Complex Temporal Relation Reasoning}} \\
\midrule
\textbf{Audio Inputs} & 
\textbf{Reference Listeners (Majority):} \texttt{[Glass Shattering $\rightarrow$ Dog Barking $\rightarrow$ Door Slamming]} \newline
\textbf{Odd Listener (Anomaly):} \texttt{[Dog Barking $\rightarrow$ Glass Shattering $\rightarrow$ Door Slamming]} \\
\midrule
\midrule

\rowcolor{boxred}
\textbf{Stage 1: Base Policy} \newline &
\textbf{Reference Clue:} "There are several loud sounds in an indoor scene, including a dog and something breaking." \newline
\textbf{Odd-Listener Clue:} "I hear a dog barking and some sharp noise in the background." \newline
\textbf{Attribution (Judge):} "Both descriptions mention a dog and a sharp breaking sound, but neither gives a reliable event order. The inconsistency is unclear. 

\textbf{Guess: Player 1 (Wrong)}." \\
\midrule

\rowcolor{boxyellow}
\textbf{Stage 2: Mid-Train} \newline &
\textbf{Reference Clue:} "A sharp glass-breaking sound occurs first, then a dog barks and a door slams near the end." \newline
\textbf{Odd-Listener Clue:} "A dog barks first, then glass shatters and a door slams."  \newline
\textbf{Attribution (Judge):} "Most descriptions place glass before the dog bark, while one description places the dog bark before the glass. This suggests a temporal mismatch. 

\textbf{Guess: Player 2 (Correct)}." \\
\midrule

\rowcolor{boxgreen}
\textbf{Stage 3: Audio-Zero} \newline &
\textbf{Reference Clue:} "`In a noisy indoor room with people talking in the background, I hear a sharp \textbf{glass shattering} sound first. Approximately \textbf{1 seconds later}, there are two quick \textbf{dog barks} and the clip ends with a heavy \textbf{door slam}." \newline
\textbf{Odd-Listener Clue (Masquerade):} "Inside a noisy indoor room, I can hear some people talking while a \textbf{sharp glass shattering}, a loud \textbf{dog bark} and a heavy \textbf{door slam} occur in close succession." \newline
\textbf{Attribution (Judge):} "Although Player 2 strategically incorporates the background context (noisy room, people talking) and matches the core event vocabulary to blend in with the majority, it completely omits the relative timeline. The most descriptions explicitly establish a strict [Glass $\rightarrow$ 1s $\rightarrow$ Dog] chronological sequence. Despite reporting identical acoustic facts, Player 2's ambiguous alignment exposes its anomaly. 

\textbf{Conclusion: Player 2 (Correct).}'' \\
\bottomrule
\end{tabular}
\caption{Visualization of the linguistic and logical self-evolution loop on a complex temporal relation sample. Under the game-theoretic pressure, the model successfully transitions from a passive semantic accumulator (Stage 1) to a precise auditory detective that organizes events into rigid causal and temporal timelines (Stage 3).}
\label{tab:case_study_table}
\end{table*}

\section{Related Work}

\paragraph{Large Audio Language Models}
LALMs connect acoustic encoders with language models to support audio-centered instruction following, reasoning, and spoken interaction.
Qwen2-Audio trains a unified audio-language model across speech, sound,
and music tasks~\citep{chu2024qwen2}, while Qwen3-Omni and Qwen3.5-Omni~\citep{xu2025qwen3,team2026qwen3} extend this line toward unified multimodal interaction with strong audio
and audio-visual capabilities. Recent models such as GLM-4-Voice and MiniCPM-o 4.5~\citep{zeng2024glm,cui2026minicpm} also explore audio foundation models for understanding, generation, and spoken interaction. These models provide strong foundations for audio reasoning, but their training or adaptation often relies on audio-text supervision, task annotations, or instruction data.
Because fine-grained auditory details are expensive to annotate systematically, existing LALMs still leave room for improvement on
benchmarks targeting temporal ordering, counting, duration, and deep
audio reasoning~\citep{kumar2025mmau,sakshi2024mmau,ma2025mmar,bhattacharya2025benchmarking}. Therefore, there is a growing need for post-training methodologies that can reduce the dependence on content-level annotations while simultaneously strengthening fine-grained auditory reasoning.

\paragraph{Post-Training and Self-Training for Audio Models}
A growing line of work aims to reduce annotation costs for audio models. SI-SDA uses self-improvement and pseudo-label filtering for unlabeled speech adaptation \citep{wang2025self}. AQA-TTRL performs test-time reinforcement learning for audio question answering using majority-vote pseudo labels \citep{zhang2025aqa}. Synthetic bootstrapping methods generate machine-authored audio-text pairs to mitigate hallucinations and augment training data without human intervention~\citep{kuan2025alignment}.
These methods reduce supervision needs, but they usually train models to imitate pseudo-labels, exploit synthetic supervision, or adapt to a specific test distribution. Audio-Zero differs by constructing a verifiable interaction environment: the model does not need to trust its own pseudo-answer as a label; success is determined by whether the hidden odd listener can be identified from generated clues.


\paragraph{Self-Play and Self-Evolution in LLMs and VLMs}
Self-play has emerged as a promising paradigm for bootstrapping language models without relying on static human annotations. 
In the text domain, frameworks rooted in SPIN~\citep{chen2024self} enable models to self-improve through iterative generation and discrimination, a philosophy rapidly extended to zero-data self-training~\citep{liu2025spiral,xu2025genius,huang2025r,zhao2026absolute} and specialized environments—such as strategic search, long-context reasoning, and multi-agent systems—where agents leverage self-guided exploration or mutual feedback to autonomously upgrade their capabilities~\citep{liu2025spice,yang2025spell,wang2025socratic}.
Similarly, vision-language approaches like Visual Self-Play and Vision-Zero~\citep{he2025visplay,wang2025vision,tong2025game} demonstrate that structured visual interactions can convert unlabeled multimodal differences into verifiable learning signals for robust decision-making.

\section{Conclusion}
In this work, we present Audio-Zero, a label-free self-evolution framework for improving fine-grained auditory perception and reasoning in large audio-language models. Instead of relying on human-annotated transcripts, QA labels, or reasoning traces, Audio-Zero converts unlabeled audio contrast pairs into an auditory self-play game, where the model alternates between generating listening-stage descriptions and performing attribution-stage reasoning to identify the odd listener. Experiments on Qwen2-Audio-7B-Instruct and Qwen2.5-Omni-7B demonstrate consistent gains on both general audio reasoning benchmarks and fine-grained temporal reasoning benchmarks. Specifically, Audio-Zero improves the two backbones by 7.33\% and 4.00\% on TREA, 7.90\% and 10.00\% on MMAR and 3.00\% and 8.00\% on MMAU Test-mini, respectively. Further analyses show that the game-derived feedback encourages the emergence of more event-rich, acoustically detailed and discriminative auditory descriptions. These results suggest that self-play over unlabeled audio contrasts is a promising direction for scalable post-training of LALMs toward stronger fine-grained auditory reasoning.

\section*{Impact Statement}
This paper presents work whose goal is to advance the field of Machine Learning. There are many potential societal consequences of our work, none which we feel must be specifically highlighted here.

\bibliography{iclr2025_conference}
\bibliographystyle{iclr2025_conference}

\onecolumn
\appendix

\section{Appendix}

\subsection{Additional Implementation Details}
\label{app:details}

\subsubsection{Training Data Construction}
We instantiate \text{Audio-Zero} using Audio-Alpaca \citep{majumder2024tango}, a paired audio preference dataset where each example contains a prompt, a chosen audio, and a rejected audio. We treat the chosen and rejected audios as an audio contrast pair $(a^{\mathrm{ref}}, a^{\mathrm{var}})$. The two clips share prompt-level semantics but may differ in event order, missing or extra events, repetitions, acoustic quality, or background scene.

We sample and filter 2{,}000 pairs from the full collection for post-training. We remove examples whose audio files cannot be loaded, whose duration is too short or too long for stable batching, or whose paired audios are near-duplicates after manual or automatic inspection. All selected clips are resampled to 16~kHz and truncated or padded to a fixed duration during training. The dataset is only used as a source of unlabeled auditory contrast; no caption, QA label, or reasoning trace is used as supervision.

\subsubsection{Game Configuration}
Unless otherwise specified, each self-play game uses $N=4$ players and $R=2$ listening rounds. For each game, one player is uniformly sampled as the odd listener. The odd listener receives $a^{\mathrm{var}}$, while the remaining players receive $a^{\mathrm{ref}}$. The model sequentially plays all player roles and generates one listening-stage clue for each player in each round. After all clues are generated, the model acts as the judge and predicts the odd listener based on the full clue history.

The odd-listener identity is never provided to the model as input. It is only used by the environment to compute the attribution reward. This ensures that the training signal is verifiable by construction while remaining independent of human-annotated content labels.

\subsubsection{Reward Implementation}
\label{app:reward}
The attribution reward is implemented as a binary correctness signal:
\begin{equation}
    R_{\mathrm{att}} = \mathbb{I}[\hat{s}=s].
\end{equation}
For listening-stage descriptions, we combine a content utility reward with a vote-aware penalty. Mathematically, the total reward allocated to player $i$ is formalized as:
\begin{equation}
    R_{\mathrm{list}}^{i} = R_{\mathrm{utility}}^{i} \;-\; \lambda \cdot \frac{v_i}{\max(1,\,V-1)}
\end{equation}
Here, $v_i$ is the number of attribution votes received by player $i$, $V = G \times N$ is the total number of voters, and $\lambda$ is the vote penalty coefficient. In our implementation, we set $\lambda=0.6$.

The content utility reward $R_{\mathrm{utility}}^{i}$ is computed with an automated rule-based scorer. It rewards clues that mention concrete auditory events, temporal relations, quantity expressions, acoustic attributes, and scene context. It penalizes overly vague descriptions, identity leakage, and repeated phrasing across rounds. The vote-aware penalty discourages clues that make a player appear suspicious, encouraging descriptions that are informative and discriminative without directly revealing the hidden role.

\subsubsection{GRPO Hyperparameters}
\label{app:hyperparams}
Table~\ref{tab:hyperparams} summarizes the main hyperparameters used in our post-training experiments.

\begin{table}[h!]
\centering
\small
\begin{tabular}{lc}
\toprule
\textbf{Hyperparameter} & \textbf{Value} \\
\midrule
Number of players $N$ & 4 \\
Listening rounds $R$ & 2 \\
GRPO completions $G$ & 6 \\
Training pairs & 2{,}000 \\
Learning rate & $1\times10^{-5}$ \\
KL coefficient $\beta_{\mathrm{KL}}$ & 0.04 \\
Vote penalty $\lambda$ & 0.6 \\
Precision & \texttt{bfloat16} \\
Optimizer / Trainer & GRPO \\
\bottomrule
\end{tabular}
\caption{Main hyperparameters for Audio-Zero post-training.}
\label{tab:hyperparams}
\end{table}

\section{Evaluation Protocol}
\label{app:evaluation}
We evaluate all models using multiple-choice question accuracy. For MMAU Test-mini, we report results on Sound, Music, and Speech subsets as well as the overall average. For TREA, we report Order, Count, and Duration subsets. For MMAR, we report Sound, Music, and Speech subsets, and also compute the overall score over the full benchmark, including mixed-modality samples.

For each question, the model is prompted to output only the option letter or the exact choice text depending on the benchmark format. Generation is performed greedily without sampling. We use the same evaluation script and answer extraction rules for all methods to ensure fair comparison.

\section{Diagnostic Metrics}
\label{app:diagnostic_metrics}
To analyze whether fine-grained auditory descriptions emerge during training, we compute five diagnostic metrics over listening-stage clues. All feature-based metrics are length-normalized where applicable and min-max scaled to $[0,1]$ for visualization in Figure~\ref{fig:diagnostic}. Table~\ref{tab:diagnostic_metrics} summarizes their rigorous matching definitions and conceptual descriptions.

\begin{table*}[t!]
\centering
\small
\setlength{\tabcolsep}{8pt}
\begin{tabular}{p{0.24\linewidth} p{0.46\linewidth} p{0.24\linewidth}}
\toprule
\textbf{Metric} & \textbf{Definition} & \textbf{Target Examples} \\
\midrule
\textit{Distinct Event Count} &
Number of unique auditory event words mentioned in a clue. It measures whether the model describes concrete sound events rather than generic audio content. &
speech, footsteps, knock, music, glass shattering \\
\addlinespace
\textit{Temporal Relation Count} &
Number of explicit ordering relations between events. It measures whether the model organizes events into temporal structure. &
after, before, followed by, then, first \\
\addlinespace
\textit{Quantity Expression Frequency} &
Frequency of quantity or repetition expressions. It measures whether the model captures counting-related auditory details. &
twice, repeated, several, multiple, two barks \\
\addlinespace
\textit{Acoustic Attribute Density} &
Density of acoustic or auditory attribute words. It measures whether the model describes sound quality and perceptual properties. &
loud, faint, muffled, distant, clear, sharp, heavy \\
\addlinespace
\textit{Vagueness Rate} &
Proportion of clues containing vague or underspecified phrases. Lower values indicate more specific descriptions. &
some sounds, various noises, something happens \\
\bottomrule
\end{tabular}
\caption{Diagnostic metrics for analyzing listening-stage descriptions.}
\label{tab:diagnostic_metrics}
\end{table*}

\section{Prompt Templates}
\label{app:prompts}
We provide the exact input templates wrapper used during training in the following structured layout.

\subsection{Listening Stage Instruction Template}
The model is instructed to describe what the current player hears without revealing its hidden role. The prompt emphasizes concrete auditory evidence, including events, order, repetitions, acoustic attributes, and background scene.

\begin{center}
\begin{minipage}{0.96\linewidth}
\begin{tcolorbox}[colback=gray!4,colframe=gray!40,arc=2mm,title=\textbf{Prompt Template: Clue Generation (Listening Round $r$)}]
\small
\texttt{[System Instruction]} \\
You are Player $i$ in an audio self-play game. Listen to your private audio and provide a concise clue describing what you hear. Mention concrete sound events, temporal order, repetitions, acoustic properties, and scene context when relevant. Do not state whether you are the odd listener. \\\\
\textbf{Input Audio:} \\
\textbf{Public Clue History:}
\end{tcolorbox}
\end{minipage}
\end{center}

\subsection{Attribution Stage Instruction Template}
The model is instructed to act as the judge and identify the odd listener from the full clue history.

\begin{center}
\begin{minipage}{0.96\linewidth}
\begin{tcolorbox}[colback=gray!4,colframe=gray!40,arc=2mm,title=\textbf{Prompt Template: Odd-Listener Identification (Judge Perspective)}]
\small
\texttt{[System Instruction]} \\
You are the judge in an audio self-play game. Most players heard the same reference audio, while one odd listener heard a subtle variant. Given the full set of player clues, compare their descriptions and identify which player most likely heard the variant audio. Return the player number. \\\\
\textbf{Reference Audio Input:} \\
\textbf{Compiled Player Clues:}
\end{tcolorbox}
\end{minipage}
\end{center}

\section{Additional Notes on Reproducibility}
\label{app:reproducibility}
All experiments are run with the same random seed for data sampling and role assignment unless otherwise specified. For checkpoint evaluation, we use the same benchmark splits and answer extraction rules across all methods. For diagnostic analysis, the same set of generated listening-stage clues is used to compute all metrics at each training iteration.

\end{document}

%% file: math_commands.tex
\usepackage{amsmath,amsfonts,bm}

\def\eqref#1{equation~\ref{#1}}

\def\1{\bm{1}}

\DeclareMathAlphabet{\mathsfit}{\encodingdefault}{\sfdefault}{m}{sl}
\SetMathAlphabet{\mathsfit}{bold}{\encodingdefault}{\sfdefault}{bx}{n}

